# Bilayer Graphene Growth via a Penetration Mechanism

*Ping Wu, Xiaofang Zhai, Zhenyu Li,\* Jinlong Yang*

Hefei National Laboratory for Physical Sciences at the Microscale and Synergetic Innovation Center of Quantum Information and Quantum Physics, University of Science and Technology of China, Hefei, Anhui, 230026, China

\*Address correspondence to zyli@ustc.edu.cn

ABSTRACT

From both fundamental and technical points of view, a precise control of the layer number of graphene samples is very important. To reach this goal, atomic scale mechanisms of multilayer graphene growth on metal surfaces should be understood. Although it is a geometrically favorable pathway to transport carbon species to interface and then form a new graphene layer there, penetration of a graphene overlayer is not a chemically straightforward process. In this study, the possibility of different active species to penetrate a graphene overlayer on Cu(111) surface is investigated based on first principles calculations. It is found that carbon atom penetration can be realized via an atom exchange process, which leads to a new graphene growth mechanism. Based on this result, a bilayer graphene growth protocol is proposed to obtain high quality samples. Such a penetration possibility also provides a great flexibility for designed growth of graphene nanostructures.

## INTRODUCTION

Graphene, a two-dimensional monolayer of $sp^2$-hybridized carbon atoms, has attracted extensive research interest due to its extraordinary physical and chemical properties with various potential applications.[1] To be used in digital electronics, however, monolayer graphene has the well known zero-gap issue[2-3] which makes the achievement of a high on-off ratio difficult. To solve this problem, bilayer graphene can be used instead, with a gap opening simply by applying an electric field.[4-5] The layer number is thus an important parameter to control in graphene sample preparation.

Currently, there are several ways to produce graphene.[6-7] However, it is still a great challenge to simultaneously obtain both good sample quality and high process scalability. Chemical vapor deposition (CVD) based graphene growth on metal surfaces is a potential method to reach this goal.[8-9] On substrates with a high carbon solubility, such as Ni, it is difficult to control the number of graphene layers.[10] Cu as a metal with very low carbon solubility is then widely used to grow high quality and large area graphene monolayer. Since C segregation and precipitation are effectively avoided,[11] graphene growth on Cu substrate is believed to be a self-limiting process.

Interestingly, with specially designed growth parameters, it is also possible to grow multiple graphene layers on the Cu surface. Recently, ambient pressure (AP) CVD[12-16] has been used as an effective way to grow multilayer graphene samples. Under low pressure, bilayer or multilayer graphene can also be obtained by slowing the cooling rate,[17] providing upstream bare Cu surface,[18-19] or changing other growth conditions.[20-22] To precisely control the layer number, it is important to understand the growth mechanisms at the atomic scale. However, without C segregation and precipitation, the mechanism of multilayer graphene growth on Cu surface is still elusive.

A possible mechanism is forming bilayer or multilayer nuclei at the initial stage of growth.[13-15] Then, the layer closer to the substrate is expected to grow much faster than the up layers. Such an on-top growth mechanism is supported by the fact that hydrogen etching can effectively remove multilayer domains.[23] Another possibility is that a new graphene layer nucleates and grows directly on Cu surface under graphene overlayers.[16] Such an underlayer growth model has also been confirmed by several recent experiments.[24-26] Since direct nucleation of multilayer domains is difficult to control, we focus on the second mechanism here.

A key integrant of the underlayer growth model is the intercalation of carbon atoms to the interface between Cu substrate and the graphene overlayer. As shown in Figure 1, there are two possible intercalation pathways: diffusion through graphene edges[16] and penetration of the graphene overlayer. The latter has several advantages in bilayer graphene growth. (1) It will not be blocked when the graphene overlayer grows very large or even covers the whole substrate. (2) It leads to a more homogeneous carbon atom distribution under the overlayer, which is desirable for the formation of a high quality new layer. (3) The growth of the second layer is independent with the first layer, which will provide more flexibility for bilayer graphene growth in a controllable way.

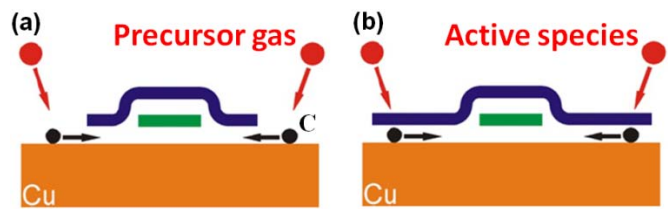

**Figure 1.** (a) Precursor gases decompose on Cu surface and the formed C atoms diffuse through edges of the graphene overlayer. (b) Some active species can penetrate a graphene overlayer, which leads to carbon intercalation and growth of the second graphene layer.

Although it is more desirable, the penetration pathway of C intercalation seems not feasible at first sight. Actually, graphene has been used as an atomic membrane impermeable to standard gases including helium.[27] Even for H atom, to pass through graphene, there is a more than 2 eV energy barrier to conquer.[28] Therefore, C atom was not expected to be able to penetrate graphene. In this article, we find that, however, C atoms can be transported to the graphene/Cu interface based on an exchange mechanism. These intercalated C atoms then provide an ideal carbon source for the second graphene layer growth. Desirably, when a graphene bilayer is formed, C atom penetration becomes prohibited, which close the way to a third layer growth. On the basis of these results, we propose a new protocol to grow high quality bilayer graphene in a controllable way. The optimal growth condition is estimated from first principles.

**RESULTS**

The energetically favorable reaction pathway for carbon atom penetration of a graphene overlayer on Cu(111) surface is shown in Figure 2a. In the initial configuration, a carbon atom adsorbs on the graphene overlayer at a bridge site, with an adsorption energy of 2.82 eV. By pushing a carbon atom in graphene to the interfacial Cu face-centered cubic (fcc) hollow site and substituting it, the carbon adatom can "penetrate" into the interface, with a significant exothermic reaction energy (2.40 eV). The energy barrier of this process is 0.93 eV. After conquering another 0.41 eV barrier, the chemical bond between the intercalated C atom and graphene is fully broken. This intercalated carbon atom finally goes to the most stable subsurface octahedron site, which is 2.90 eV more favorable in energy compared to the bridge site on graphene. Therefore, there is a big thermodynamic driving force moving carbon adatoms on graphene to Cu subsurface sites via a penetration or more strictly speaking exchange mechanism.

uncovered metal surfaces which can diffuse with a walk-with-legs model.[30] Therefore, nucleation at the interface are expected to be more difficult than on uncovered Cu(111) surface. This is actually desirable for graphene growth, since lower nucleation densities leads to larger single-crystal domain sizes.

Once a nucleus is formed, the second graphene layer starts to grow. For simplicity, a compact $C_6$ ring is used to mimic the edge of a graphene island.[31] Considering the relatively low diffusion barrier of C adatoms on the graphene overlayer (0.68 eV, see Figure S2), we first check the possibility of direct penetration and attachment at the edge of an underlying graphene island. The energy barrier associated with this process is 1.11 eV as shown in Figure 3a. Since the concentration of edge sites is low, this process will not make an important contribution to the second graphene layer growth. A carbon atom can also attach to $C_6$ from a subsurface octahedron site by conquering a 1.08 eV barrier (Figure 3b). With a graphene overlayer on the surface, carbon atom diffusion in the Cu subsurface layer associates a barrier only 0.30 eV, much smaller than this attachment barrier. Therefore, the growth of the second layer is an attachment limited process.

**Figure 3.** Minimum energy paths of C monomer incorporation in the second graphene layer growth from (a) a bridge site on graphene and (b) a subsurface octahedron site.

Since a uniform bilayer graphene is very desirable, it is important to check if the same penetration mechanism will lead to multilayer graphene growth. For this purpose, we check the possibility of a carbon atom penetrating an already formed graphene bilayer (Figure 2b). As the first step, passing the top layer via atom exchange has an energy barrier of 1.48 eV, which is already significantly higher than that of penetrating a single layer graphene. After passing through the first top layer, penetrating the second layer has an energy barrier as high as 2.94 eV. Therefore, penetration of a graphene bilayer is much more difficult than a single graphene overlayer, which makes a controllable growth of uniform bilayer graphene possible.

Hydrogen has also been widely used in graphene growth on metal surface.[8-9] It is thus interesting to check its effect on penetration based bilayer graphene growth. When H is contained in an active species supplied for the second graphene layer growth, the penetration process is expected to become more difficult. As shown in Figure 4, although dissociative penetration of a graphene overlayer by the carbon atom in a CH radical is still exothermic (notice that decomposition of CH on bare Cu surface is thermodynamically not preferred[32]), the energetic driving force decreases to 0.91 eV.

**Figure 4.** Penetration pathway of a CH radical on the graphene/Cu (111) surface.

To penetrate the graphene overlayer, the CH radical need to move from the bridge site to a top site first, then it can push the carbon atom below to the interface. The energy barrier associated with these two processes is 1.08 and 0.75 eV, respectively. By conquering another 1.07 eV barrier, the final stable state can be reached, where the interfacial carbon is still bonded with graphene. Even if the penetrated carbon has already diffused to a subsurface site, H adsorbed on graphene nearby can still drag it up to relax strain on graphene (the green path in Figure 4). On the basis of these results, hydrogen seems not helpful at the stage of the second layer growth. In fact, for $CH_2$ and $CH_3$, penetration even becomes endothermic with an energy penalty of about 1.30 and 3.03 eV (Figures S3 and S4), respectively.

Although it is not required during the second layer growth, hydrogen may be useful at the final stage when there are some extra carbon species on the graphene overlayer to be etched away. To study this effect, we first check the possibility of carbon aggregation on the graphene overlayer.

As an example, we consider the combination of two carbon adatoms before their penetration happens, which leads to the formation of a $C_2$ cluster and the generation of a 5775 defect on the graphene (Figure 5). The overall energy barrier associated to such a $C_2$ dimer formation is about 0.46 eV (Figure S5). We note that perpendicular carbon chain has been suggested to be a stable adsorption configuration of carbon clusters on graphene.[33] Our test calculations indicate that even on freestanding graphene if a large supercell (5×5, for example) is used, the 5775 defect is still more stable than vertical $C_2$ cluster.

In the 5775 topological defect, carbon dimer prefers to be located slightly lower than the graphene plane, and the energy barrier from the $C_2$-above to $C_2$-below configuration is relatively small (0.26 eV). However, a complete $C_2$ penetration turns out to be very difficult with a large energy barrier of 2.29 eV. Therefore, combination of two carbon adatoms will generate extra carbon and leave defect on the graphene overlayer. When hydrogen is provided, it can adsorb on the dimer. After conquering a 1.41 eV barrier, desorption of an acetylene molecule finally heals the graphene overlayer (Figure S6).

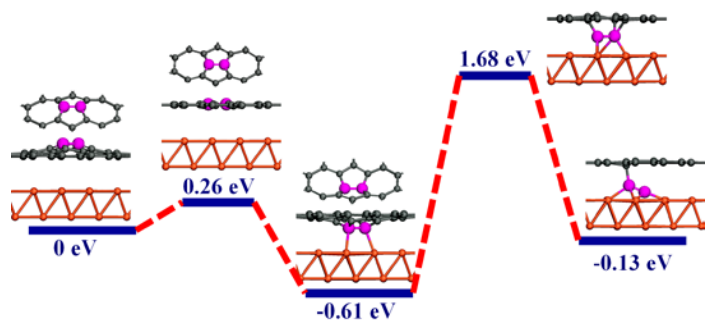

**Figure 5.** Minimum energy path of carbon dimer penetration of a graphene overlayer on Cu surface.

On the basis of the above results, a layer-by-layer growth of bilayer graphene is possible. To obtain high quality bilayer graphene, we propose a three-step protocol as illustrated in Figure 6. The main idea is using relatively well developed techniques to grow the first graphene layer, then growing the second layer based on the penetration mechanism, and finally using hydrogen to clean the top graphene layer which may have been contaminated during the second step. Since the atomic mechanism is clear, optimal growth parameters can be estimated from first principles.

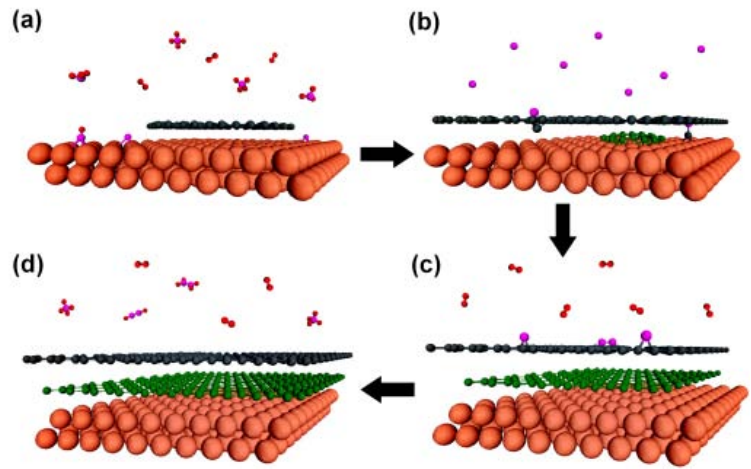

**Figure 6.** Schematic of a new bilayer graphene growth protocol. (a) A standard monolayer graphene CVD growth on Cu surface. (b) Carbon monomer intercalation via a penetration mechanism and growth of the second graphene layer. (c) Hydrogen gas is supplied to etch extra carbon species. (d) With desorption of hydrocarbons, high quality bilayer graphene is obtained.

The first step is growing a high quality large area single layer graphene. This is a widely studied topic, and there are already some recipes in the literature to reach this goal, both experimental demonstrations[34-35] and theoretical predictions.[36] In principle, by controlling the nucleation behavior, it is possible to grow large area graphene single crystal. At this step, a 100% graphene coverage is desirable to avoid carbon atom intercalation from graphene edges.

In conventional CVD growth, it will be difficult to effectively transport active carbon species to the top of graphene overlayer. Therefore, at the second step, carbon monomer is directly provided, possibly from an effusion cell with a heated pyrolytic graphite filament.[37] With careful control of the heating condition and in-situ monitoring of the atomic flux by the available molecular beam epitaxy (MBE) technique,[38] the carbon flux can be precisely controlled. The maximum carbon flux allowed to avoid amorphous carbon aggregation on the surface of graphene can be estimated as:

$$F^{max} = \frac{k}{S} = \frac{A}{S} e^{-E_a/k_B T}$$

where $k$ is the penetration rate, which can be calculated using the Arrhenius equation, with the calculated activation energy $E_a$ (0.93 eV) and a typical estimation ($10^{13}$ s$^{-1}$) of the pre-exponential factor $A$.[39] $S$ is the surface area of an adatom capture zone, beyond which carbon cluster formation before penetration can be avoid. Both $k$ and $S$ are temperature dependent.

Before giving a value for $F^{max}$, we first estimate the upper bound of the growth temperature at this step ($T^{max}$) by considering the requirement to effectively depress the growth of the third layer. Notice that the carbon atom penetration barrier for a single graphene overlayer ($E_1$) is 0.93 eV. After the bilayer is formed, energy barrier for the penetration of the first layer ($E_2$) increases to 1.48 eV. If we let the ratio of carbon penetration of graphene layers with and without the second graphene layer to be smaller than $10^{-4}$, the possibility of carbon atoms entering the space between the first and second graphene layers will be practically negligible. Then $T^{max}$ can be simply estimated by the following equation

$$e^{-(E_2-E_1)/k_B T^{\max}} = 10^{-4},$$

which gives a $T^{\max}$ of 693 K. At this temperature, the maximum number of graphene layers formed via the penetration mechanism is two. Since 300 °C is already high enough for MBE growth of graphene,[37] this temperature is also expected to be able to give a satisfactory growth rate.

At 693 K, penetration rate $k$ is $1.72 \times 10^6$ s$^{-1}$. Since the diffusion barrier of carbon adatom on a graphene overlayer ($E_d$, 0.68 eV) is lower than the penetration barrier ($E_p$, 0.93 eV). A carbon adatom will diffuse a long distance before penetration. The ratio of diffusion and penetration probabilities can be estimated as:

$$P = \frac{e^{-E_d/k_B T}}{e^{-E_p/k_B T}}$$

At 693 K, $P$ is about 65.8, which means that the penetration probability at each diffusion step is about 0.015. After 610 diffusion steps, the probability that a carbon penetration still does not happen will be smaller than $10^{-4}$. Therefore, it is safe to set the adatom capture zone as a 305×305 graphene supercell, since two diffusion steps match a lattice parameter distance. The corresponding area S then equals to $5.32 \times 10^5$ Å$^2$, and $F^{\max}$ is as large as $3.23 \times 10^{16}$ cm$^{-2}$s$^{-1}$. This value is five orders of magnitude larger than a typical carbon flux number in MBE experiment ($10^{11}$ cm$^{-2}$s$^{-1}$).[37] As a result, in our protocol, aggregation of carbon adatoms on the graphene overlayer can be effectively suppressed.

With well controlled carbon flux, concentration of extra carbon species on the graphene will be very low. Therefore, large carbon aggregates will not be formed. For small carbon clusters like dimer, we have demonstrated that they can be effectively etched away by hydrogen. Therefore, the third step is introducing hydrogen gas to etch the remaining carbon species and heal defects. After this step, a high quality graphene bilayer sample is expected to be obtained.

**DISCUSSION**

An important merit of our protocol is that, since the first layer can be considered as a template for the second layer growth, good stacking order can be expected in our bilayer graphene samples. At the same time, the protocol proposed here provides a great flexibility for designed growth. For example, with a patterned substrate,[36] we can induce patterned heterogeneous nucleation. Then interesting graphene bilayer pattern with potential electronics applications can be obtained straightforwardly, since now the growth of the second layer is totally decoupled with that of the first layer and it can be terminated at any time. Also notice that active species, such as CH, can exist in gas phase in conventional CVD growth of graphene.[20] Therefore, the penetration mechanism may have also played an important role in previous graphene growth experiments.

Cu(100) surface is another important surface for practical graphene growth on Cu foil. In test calculations using graphene nanoribbon as a model system of the graphene overlayer (Figure S7), we obtain a carbon atom penetration barrier of 0.75 eV, even lower than that in the Cu(111) case. This result can be understood by the stronger carbon-surface interaction on the Cu(100) surface. Consistently, carbon penetration on the Cu(100) surface lower the total energy by 4.11 eV, which is more favorable than on the Cu(111) surface. Since Cu is located at the low end of metal-carbon interaction strengths, the exchange based penetration mechanism is expected to be quite universal for typical metal substrate used in graphene growth.

**CONCLUSIONS**

In summary, the penetration mechanism of bilayer graphene growth has been revealed for the first time, a carbon exchange process makes this otherwise unlikely penetration pathway possible. Fortunately, further penetration of an already formed graphene bilayer is very difficult. Therefore, penetration provides an ideal pathway to transport carbon atoms for the second layer growth under the first layer. With these inspiring insights, a bilayer graphene growth protocol with great controllability and flexibility has been proposed. Optimal growth parameters have been estimated from first principles. The penetration mechanism reported here is expected to open a new avenue in graphene synthesis chemistry.

METHOD

Density functional theory (DFT) calculations were performed with the Vienna ab initio simulation package (VASP),[40-41] using the Perdew−Burke−Ernzerhof (PBE) exchange-correlation functional.[42] To describe the van der Waals (vdW) interaction, the Grimme's DFT-D2 method[43] was adopted. A four-layer slab models were used to describe the Cu (111) surfaces, in which the bottom layer was fixed to the optimized bulk geometry with a Cu−Cu bond length of 2.57 Å. The repeated slabs were separated by more than 10 Å to avoid interactions between neighboring slabs. A 400 eV kinetic energy cutoff was chosen for plane-wave basis set, and the Monhorst-Pack $k$-point sampling[44] parameters were carefully tested to produce well-converged results. Graphene lattice parameter was slightly adjusted to match that of the Cu (111) surface. The interfacial distance between the graphene and the substrate is calculated to be 3.0 Å, agreeing with previous theoretical studies.[45] The most stable configuration of graphene on Cu substrate has its hexagons centered at hexagonal close packed (hcp) hollow sites (Figure S8), as also reported previously.[7,46] The climbing image nudged elastic band (CI-NEB) method [47] was used for transition state search and barrier height determination. Residual forces were within 0.02 eV/Å for geometry optimizations and 0.03 eV/Å for transition state location.

ACKNOWLEDGMENT

This work is partially supported by MOST (2011CB921404, 2014CB932700), by NSFC (21173202, 21222304, and 21121003), by CUSF, by CAS (XDB01020300), and by USTC-SCC, SCCAS, and Tianjin Supercomputer Centers.

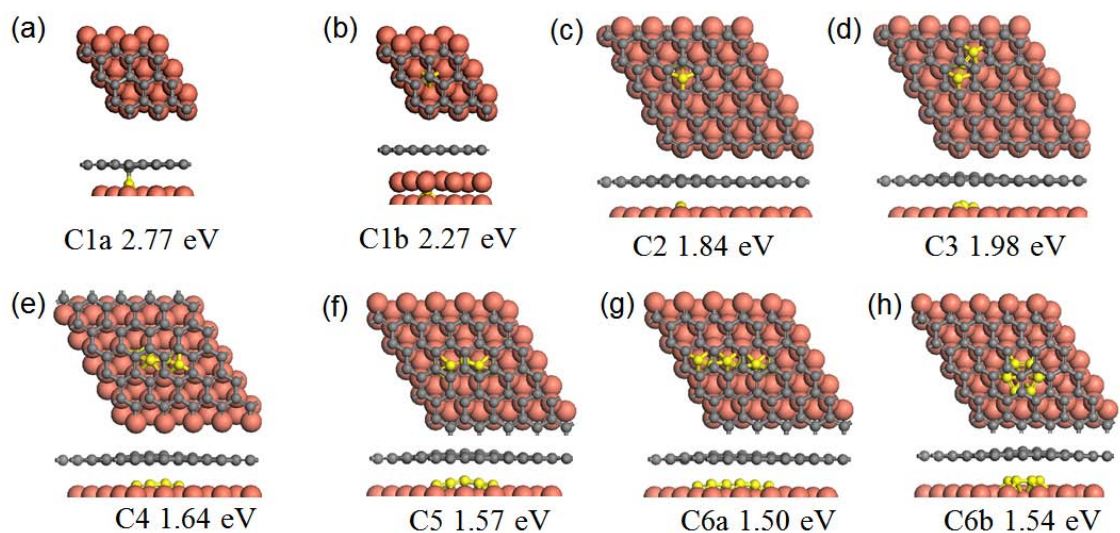

**Figure S1.** Stable structures (N=1-6) of carbon clusters in the interface between graphene and Cu(111) substrate. The marked potential energy is generally defined as $(E_{tot} - E_{G/Cu(111)} - N_C \times E_C - N_H \times E_H)/N_C$, where $E_{tot}$ and $E_{G/Cu(111)}$ are energies of the carbon species adsorbed system and the clean Gaphene/Cu surface. $E_C$ is the energy of a carbon atom in graphene, and $E_H$ is the energy of a hydrogen atom in the $H_2$ molecule. $N_C$ and $N_H$ are the numbers of carbon atoms and hydrogen atoms in the adsorbed carbon species, respectively. For carbon clusters shown in this figure, $N_H = 0$.

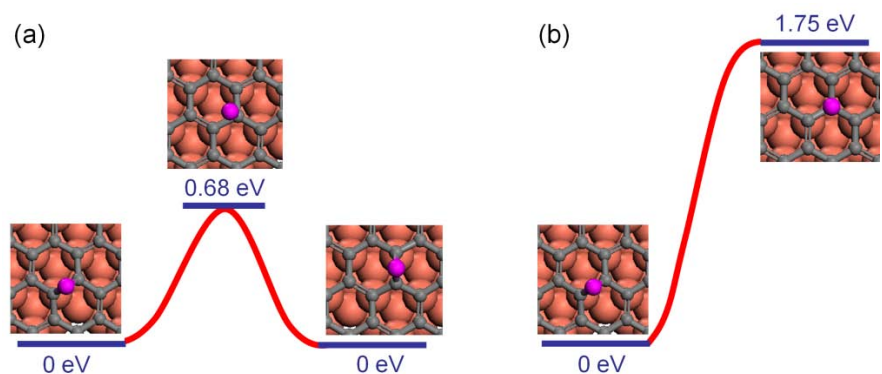

**Figure S2.** Minimum energy paths of C monomer diffusion on graphene/Cu(111) (a) between two bridge sites and (b) from bridge to top sites.

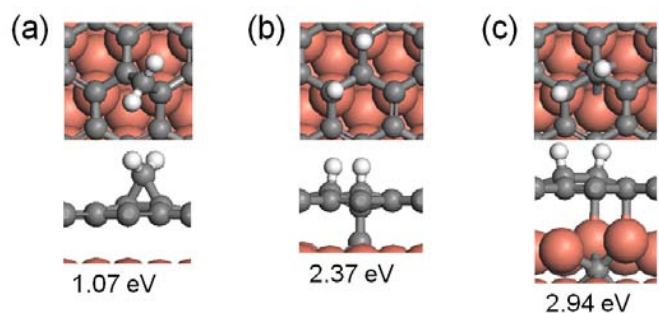

**Figure S3.** Top and side views of (a) a $CH_2$ adsorbed on graphene/Cu(111), (b) the carbon atom in the interface, and (c) the carbon atom at a subsurface octahedron site. Potential energies of $CH_2$ as defined in Figure S1 are marked in eV.

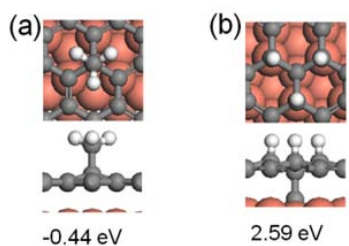

**Figure S4.** Top and side views of (a) a $CH_3$ adsorbed on graphene/Cu(111), (b) the carbon atom in the interface. Potential energies of $CH_3$ as defined in Figure S1 are marked in eV.

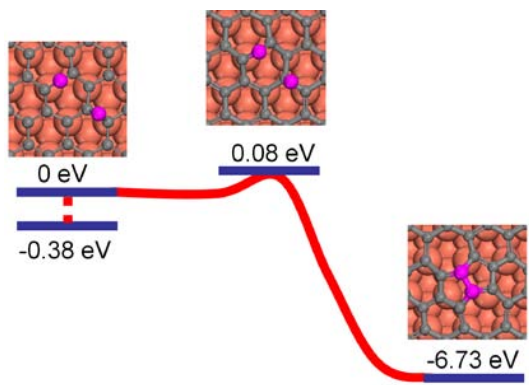

**Figure S5.** Minimum energy path of carbon monomes collision with each other to form dimer embedding in the graphene overlayer. The initial state is 0.38 eV higher than that of two carbon adatoms far away each other.